\documentclass{ws-procs9x6}

\begin{document}

\title{Dependence of Nuclear Binding Energies
on the Cutoff Momentum of Low-Momentum Nucleon-Nucleon Interaction}

\author{S. Fujii}

\address{Department of Physics, University of Tokyo, \\
Tokyo 113-0033, Japan\\
E-mail: sfujii@nt.phys.s.u-tokyo.ac.jp}

\author{H. Kamada, R. Okamoto and K. Suzuki}

\address{Department of Physics, Kyushu Institute of Technology, \\ 
Kitakyushu 804-8550, Japan}

\maketitle

\abstracts{
Binding energies of $^{3}$H, $^{4}$He, and $^{16}$O are calculated,
using low-momentum nucleon-nucleon interactions ($V_{{\rm low}-k}$)
for a wide range of the cutoff momentum $\Lambda$.
In addition, single-particle energies in nuclei around $^{16}$O are computed.
The dependence of the binding energies and the single-particle energies
in these nuclei on the cutoff momentum $\Lambda$ of the $V_{{\rm low}-k}$
is examined.
Furthermore, the availability of the $V_{{\rm low}-k}$
in nuclear structure calculations is discussed.
}

\section{Introduction}
One of the fundamental objectives in nuclear structure calculations is
to describe nuclear properties,
starting with high-precision nucleon-nucleon interactions.
Since this kind of interaction has a repulsive core at a short distance,
one has been forced to derive an effective interaction or $G$ matrix
in a model space for each nucleus from the realistic interaction,
except for precise few-nucleon structure calculations.

Recently, Bogner {\it et al.} have constructed low-momentum nucleon-nucleon
interactions $V_{{\rm low}-k}$ from high-precision nucleon-nucleon
interactions to use them as microscopic input to the nuclear many-body
problem.\cite{Bogner03}
The $V_{{\rm low}-k}$ can be derived using techniques of conventional
effective interaction theory or renormalization group method.
They have shown that the $V_{{\rm low}-k}$ conserves the properties
of the original interaction,
such as the half-on-shell $T$ matrix and the phase shift within
a cutoff momentum $\Lambda$ which specify the low-momentum region.
The $V_{{\rm low}-k}$ for the typical cutoff $\Lambda=2.1$ fm$^{-1}$
corresponding to $E_{\rm lab}\simeq 350$ MeV are almost the same and
are not dependent on the realistic nucleon-nucleon interactions employed.
Thus, as a $unique$ low-momentum interaction,
the $V_{{\rm low}-k}$ at approximately $\Lambda \sim 2$ fm$^{-1}$
has been employed directly in nuclear structure calculations,
such as the shell-model\cite{Bogner02} and the Hartree-Fock
calculations.\cite{Coraggio03}
Especially, the calculated excitation spectra in the shell-model calculations
show the good agreement with the experimental data
and are even better than those using the sophisticated $G$ matrix.
Thus, the application of $V_{{\rm low}-k}$ to nuclear structure calculations
has been growing.
We should notice, however, that the $V_{{\rm low}-k}$ is derived introducing
the cutoff momentum $\Lambda$,
and thus the calculated results using the $V_{{\rm low}-k}$
have the $\Lambda$ dependence to some extent.

One of the central aims of the present work is to examine
the $\Lambda$ dependence in structure calculations.
First, we calculate binding energies for few-nucleon
systems for which precise calculations can be performed,
and confirm the validity of the $V_{{\rm low}-k}$ in
the structure calculation by comparing the obtained results
with the $exact$ values.
Second, we proceed to heavier systems such as $^{16}$O and investigate
not only the total binding energy itself but also the single-particle energy
which is defined as the relative energy of neighboring two nuclei
such as $^{16}$O and $^{15}$O.
Through the obtained results,
we discuss the applicability of the $V_{{\rm low}-k}$ to
nuclear structure calculations.

\section{Results and discussion}

In the following structure calculations, we use the $V_{{\rm low}-k}$
which is derived from the CD-Bonn potential\cite{Machleidt96}
by means of a unitary transformation theory.\cite{Okubo54,Suzuki94}
The details of deriving the $V_{{\rm low}-k}$ and its numerical accuracy can be
seen in Ref.~\refcite{Fujii-low}.

\subsection{$^{3}$H and $^{4}$He}

In order to investigate the sensitivity of $\Lambda$ to the binding energies
of $^{3}$H and $^{4}$He precisely,
we have performed the Faddeev and the Yakubovsky calculations,
respectively.\footnote{The collaboration with E.~Epelbaum and W.~Gl\"ockle
in this part of the present work which has already been done
in Ref.~\refcite{Fujii-low} is highly appreciated.}
For simplicity,
only the neutron-proton interaction is used for all the channels.

\begin{figure}[ht]
\centerline{\epsfxsize=11cm\epsfbox{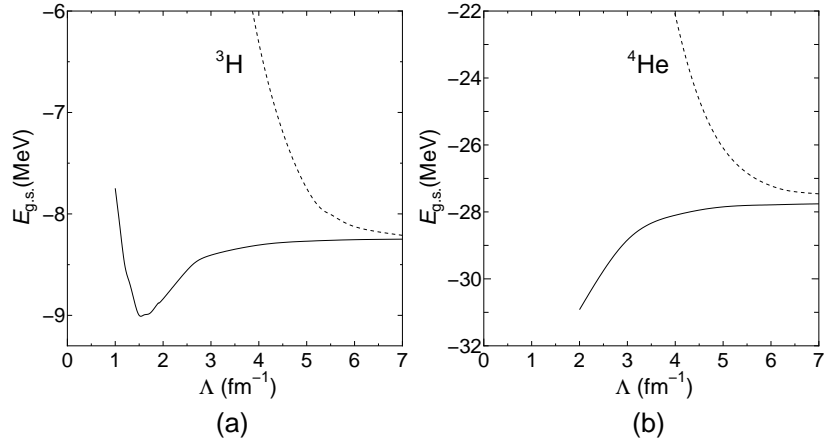}}   
\caption{Calculated ground-state energies of $^{3}$H (a) and $^{4}$He (b)
as a function of the cutoff momentum $\Lambda$.
The solid lines represent the results using the $V_{{\rm low}-k}$
for each $\Lambda$.
The short-dashed lines are the results using the original CD-Bonn potential,
where the high-momentum components beyond $\Lambda$ are simply truncated
in the structure calculation.
\label{H3-He4}}
\end{figure}

Figure~\ref{H3-He4}(a) exhibits the calculated ground-state energies
of $^{3}$H by a 34-channel Faddeev calculation
as a function of the cutoff momentum $\Lambda$.
The $exact$ value using the original CD-Bonn potential
on the above assumptions is $-8.25$ MeV.
The solid line depicts the results using the $V_{{\rm low}-k}$
from the CD-Bonn potential.
The short-dashed line represents the results
using the original CD-Bonn potential,
where the high-momentum components beyond $\Lambda$ are simply truncated
in the structure calculation.
For the case of the original CD-Bonn potential,
we need $\Lambda \ge 8$ fm$^{-1}$
to reach the $exact$ value if the accuracy of $100$ keV is required.
This situation is largely improved if we use the $V_{{\rm low}-k}$.
Even if we require the accuracy of $1$ keV, we do not need
the high-momentum components beyond $\Lambda \sim 8$ fm$^{-1}$.
However, it should be noted that the results using the $V_{{\rm low}-k}$
for the values smaller than $\Lambda \sim 5$ fm$^{-1}$ vary considerably,
and there occurs the energy minimum at around $\Lambda =1.5$ fm$^{-1}$.
The magnitude of the difference between the $exact$ value and the calculated
result using the $V_{{\rm low}-k}$ for the representative cutoff value
$\Lambda =2.0$ fm$^{-1}$ is about $600$ keV for $^{3}$H.

A similar tendency can also be seen in the case of $^{4}$He.
We have performed the $S$-wave (5+5-channel) Yakubovsky calculation
for $^{4}$He without the Coulomb interaction.
The $exact$ ground-state energy using the original CD-Bonn potential
on the above assumptions is $-27.74$ MeV.
In Fig.~\ref{H3-He4}(b), the calculated results for $^{4}$He are shown.
The shape of the energy curve is similar to that for $^{3}$H within the region
$\Lambda \ge 2$ fm$^{-1}$.
The calculated results become more overbound as
the value of $\Lambda$ becomes smaller.
The magnitude of the difference between the $exact$ value and the calculated
result using the $V_{{\rm low}-k}$ for $\Lambda =2.0$ fm$^{-1}$
is about $3$ MeV for $^{4}$He.
This amounts to five times larger than the result of $^{3}$H.
In the case of $^{4}$He, the results for $\Lambda < 2.0$ fm$^{-1}$
are not shown due to the numerical instability in the structure calculation.

Concerning the investigation of the $\Lambda$ dependence of the ground-state
energies of the few-nucleon systems, a detailed study with three-nucleon
forces has recently been reported by Nogga {\it et al}.\cite{Nogga04}

\subsection{$^{16}$O}

In order to examine the $\Lambda$ dependence in heavier systems,
we calculate the ground-state energy of $^{16}$O within the
framework of the unitary-model-operator approach (UMOA).\cite{Suzuki94}
The details of recent calculated results for $^{16}$O and
its neighboring nuclei using modern nucleon-nucleon interactions
can be seen in Ref.~\refcite{Fujii04}.
In the present study, we follow the same calculation method in that work
except for the determination method of the harmonic-oscillator energy
$\hbar \Omega$ and the size of the model space.
In Ref.~\refcite{Fujii04},
we have searched for the optimal value of $\hbar \Omega$
that leads to the energy minimum point by investigating the $\hbar \Omega$
dependence of the ground-state energy for each modern nucleon-nucleon
interaction.
Then, we have found that the optimal values are at around $14$ MeV of which
values are very close to the value determined by empirical formula such as
$\hbar \Omega = 45A^{-1/3}-25A^{-2/3}$ MeV.
Since the optimal value was $\hbar \Omega=15$ MeV for the CD-Bonn potential,
we use this value in this work for each $V_{{\rm low}-k}$.
Furthermore, we employ the same size of the optimal model space
which is specified by the quantity $\rho _{1}$ as
$\rho_{1}=2n_{a}+l_{a}+2n_{b}+l_{b}=12$,
where $\{n_{a},l_{a}\}$ and $\{n_{b},l_{b}\}$ are the sets of
harmonic-oscillator quantum numbers for two-body states.
We note that these values of $\hbar \Omega$ and $\rho _{1}$
are not necessarily the optimal ones for each $V_{{\rm low}-k}$
in the present study.

\begin{figure}[ht]
\centerline{\epsfxsize=6cm\epsfbox{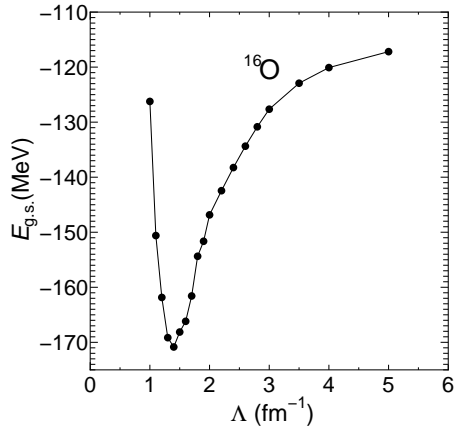}}   
\caption{The $\Lambda$ dependence of the ground-state energy of $^{16}$O using
the $V_{{\rm low}-k}$ from the CD-Bonn potential.
\label{O16}}
\end{figure}

In Fig.~\ref{O16}, the $\Lambda$ dependence using the $V_{{\rm low}-k}$
from the CD-Bonn potential of the ground-state energy of $^{16}$O is shown.
We have used the neutron-neutron, neutron-proton, and proton-proton interaction
of the CD-Bonn potential correctly for the corresponding channels,
and included the Coulomb interaction.
The partial waves up to $J=6$ are taken into account in the calculation.
The value of the ground-state energy of $^{16}$O in the $full$ calculation
given in Ref.~\refcite{Fujii04} using the original CD-Bonn potential
is $-115.61$ MeV.
Thus, the result for $\Lambda =5.0$ fm$^{-1}$ almost reproduces this value.
The calculated energy curve shows a similar tendency
to the results of $^{3}$H and $^{4}$He, but the magnitude of the difference
between the result of the $full$ calculation and the value at the energy
minimum point is considerably larger than those for $^{3}$H and $^{4}$He
due to the large difference of the mass number.
The magnitude of the difference in $^{16}$O amounts to $55$ MeV.
Even if we choose the typical cutoff $\Lambda =2.0$ fm$^{-1}$, we still observe
the significant overbinding of which magnitude is about $31$ MeV.
Thus, we may conclude from the results of $^{3}$H, $^{4}$He, and $^{16}$O
that the $V_{{\rm low}-k}$ for the typical cutoff momentum
$\Lambda \sim 2$ fm$^{-1}$ cannot reproduce
the $exact$ values, showing the significant overbinding.

It should be noted, however, that this does not necessarily mean that
the $V_{{\rm low}-k}$ for $\Lambda \sim 2$ fm$^{-1}$ is no longer valid
in nuclear structure calculations.
In fact, the shell-model calculations have shown that the $V_{{\rm low}-k}$
for $\Lambda \sim 2$ fm$^{-1}$ can work
as well as the $G$ matrix.\cite{Bogner02}

\subsection{$^{15}$O and $^{17}$O}

\begin{figure}[ht]
\centerline{\epsfxsize=11cm\epsfbox{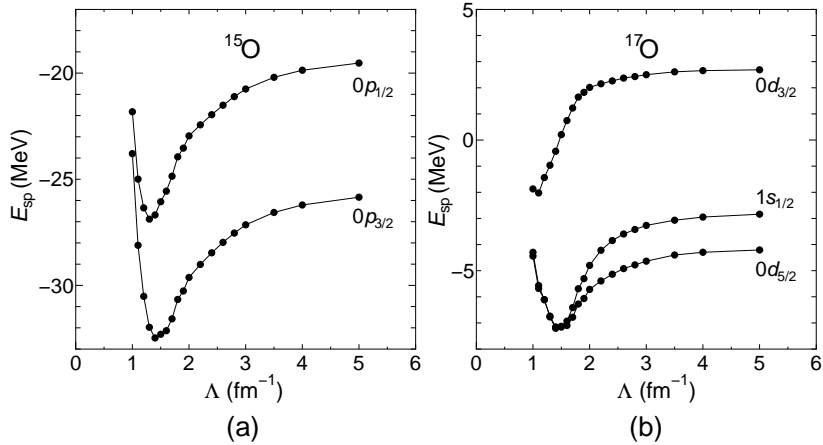}}
\caption{The $\Lambda$ dependence of the single-particle energies of
the neutron for the 0$p$ hole states in $^{16}$O which correspond to the energy
levels in $^{15}$O (a) and of the neutron particle states in $^{17}$O (b)
using the $V_{{\rm low}-k}$ from the CD-Bonn potential.
\label{O15-O17}}
\end{figure}

In the previous sections,
we have seen the results of the total binding energies.
We here examine the $\Lambda$ dependence of single-particle energies
of the neutron for hole states in $^{16}$O which correspond to the energy levels
in $^{15}$O and of neutron particle states in $^{17}$O.
The calculation procedure is essentially the same as in Ref.~\refcite{Fujii04}.
In the present study, however, we do not search for the optimal values
of $\hbar \Omega$ and $\rho _{1}$ for each single-hole or -particle state
for simplicity as in the case of $^{16}$O.
In the following calculations, we use the values of $\hbar \Omega=15$ MeV
and $\rho _{1}=12$
which are the same as in the calculation of $^{16}$O in the previous section.

Figure~\ref{O15-O17}(a) shows the $\Lambda$ dependence of
the calculated single-particle energies of the neutron for the 0$p$ hole states
in $^{16}$O which correspond to the single-hole energy levels in $^{15}$O.
The values of the $full$ calculation of the single-particle energy
are $-19.34$ and $-25.37$ MeV for the 0$p_{1/2}$ and 0$p_{3/2}$ states,
respectively.
Though the present results for $\Lambda =5.0$ fm$^{-1}$ are fairly close to
these values, there remain some discrepancies.
These discrepancies may be due to the fact that we do not search for
the optimal value of $\hbar \Omega$ for each state in the present study.
The search for the optimal values of $\hbar \Omega$ and also $\rho _{1}$
in the structure calculation with the $V_{{\rm low}-k}$ should be done
for completeness in future.

It is seen from Fig.~\ref{O15-O17}(a) that the single-particle energies for
the 0$p$ states become more attractive as the $\Lambda$ becomes smaller
as in the results of the ground-state energies.
However, what is interesting here is that the magnitudes of the spacing
between the single-particle levels, namely the spin-orbit splitting,
hold their values up to $\Lambda \sim 2$ fm$^{-1}$, although the structure of
the single-particle levels is broken within the area $\Lambda < 2$ fm$^{-1}$.

A similar tendency can also be observed in the results of $^{17}$O.
In Fig.~\ref{O15-O17}(b), the calculated results of the single-particle
energies of the neutron for the 1$s$ and 0$d$ states in $^{17}$O are shown.
The values of the $full$ calculation of the single-particle energy
are $2.67$, $-2.76$, and $-4.11$ MeV for the 0$d_{3/2}$, 1$s_{1/2}$,
and 0$d_{5/2}$ states, respectively.
The present results for $\Lambda =5.0$ fm$^{-1}$ are not so different
from these values.
The tendency of the $\Lambda$ dependence is essentially the same as
in $^{15}$O.
It can be seen again that the magnitudes of the spacings between
the single-particle levels do not vary very much within the region
$\Lambda \ge 2$ fm$^{-1}$,
while those are considerably broken within the area $\Lambda < 2$ fm$^{-1}$.
These results may suggest that the $V_{{\rm low}-k}$ for
$\Lambda \sim 2$ fm$^{-1}$ is valid
as far as relative energies from a state such as the ground state are
concerned.

\section{Conclusions}

We investigated the dependence of the ground-state energies of $^{3}$H,
$^{4}$He, and $^{16}$O on the cutoff momentum $\Lambda$ of the low-momentum
nucleon-nucleon interaction $V_{{\rm low}-k}$.
In all the cases, there appear the energy minima
at around $\Lambda=1.5$ fm$^{-1}$.
We have found that the $V_{{\rm low}-k}$ for the typical cutoff momentum
$\Lambda \sim 2$ fm$^{-1}$ cannot reproduce the $exact$ values
for the original interaction, showing the significant overbinding.
If we try to reproduce the $exact$ values, we need $\Lambda \ge 5$ fm$^{-1}$.
On the other hand, the magnitudes of the spacings between the single-particle
levels in nuclei around $^{16}$O do not so vary
within the region $\Lambda \ge 2$ fm$^{-1}$.
This may suggest that the $V_{{\rm low}-k}$ for the typical cutoff
$\Lambda \sim 2$ fm$^{-1}$ is valid in nuclear structure calculations
as far as relative energies from a state such as the ground state are
concerned as in the shell-model calculation.

\section*{Acknowledgments}
This work was supported by a
Grant-in-Aid for Scientific Research (C)
(Grant No. 15540280) from Japan Society for the Promotion of Science and
a Grant-in-Aid for Specially Promoted Research (Grant No. 13002001)
from the Ministry of Education, Culture, Sports, Science and Technology
in Japan.


\begin{thebibliography}{9}


\bibitem{Bogner03}
S. K. Bogner, T. T. S. Kuo and A. Schwenk,
{\it Phys. Rep.} {\bf 386}, 1 (2003).

\bibitem{Bogner02}
Scott Bogner, T. T. S. Kuo, L. Coraggio, A. Covello and N. Itaco,
{\it Phys. Rev.} {\bf C65}, 051301(R) (2002).

\bibitem{Coraggio03}
L. Coraggio, N. Itaco, A. Covello, A. Gargano and T. T. S. Kuo,
{\it Phys. Rev.} {\bf C68}, 034320 (2003).

\bibitem{Machleidt96}
R. Machleidt, F. Sammarruca and Y. Song,
{\it Phys. Rev.} {\bf C53}, R1483 (1996).

\bibitem{Okubo54}
S. \=Okubo, {\it Prog. Theor. Phys.} {\bf 12}, 603 (1954).

\bibitem{Suzuki94}
K. Suzuki and R. Okamoto, {\it Prog. Theor. Phys.} {\bf 92}, 1045 (1994).

\bibitem{Fujii-low}
S. Fujii, E. Epelbaum, H. Kamada, R. Okamoto, K. Suzuki and W. Gl\"ockle,
{\it nucl-th}/0404049.

\bibitem{Nogga04}
A. Nogga, S. K. Bogner and A Schwenk, {\it nucl-th}/0405016.

\bibitem{Fujii04}
S. Fujii, R. Okamoto and K. Suzuki, 
{\it Phys. Rev.} {\bf C69}, 034328 (2004).

\end{thebibliography}
\end{document}